\documentclass[twocolumn]{revtex4-2}

\usepackage[T1]{fontenc}
\usepackage[utf8]{inputenc}
\usepackage{amsmath,amssymb,bm}
\usepackage{graphicx}
\usepackage{booktabs}
\usepackage{placeins}
\usepackage{float}
\usepackage{capt-of}
\usepackage{subcaption}
\usepackage[table,xcdraw]{xcolor}
\usepackage[breaklinks=true,colorlinks=true,urlcolor=blue,
citecolor=blue,linkcolor=blue,bookmarks=false]{hyperref}

\begin{document}

\title{Gatekeepers and Hallucinations: A Layered Evaluation Framework\\
for LLM-Driven Quantum Circuit Generation}

\author{Christopher Coleman$^{1,2}$}
\email{chris.coleman@ubc.ca}
\affiliation{Department of Physics and Astronomy \& Stewart Blusson Quantum Matter
Institute, University of British Columbia, Vancouver, BC, Canada}
\affiliation{Resonance Alliance, Toronto, ON, Canada}

% REV: removed stray extra closing brace after the affiliation below
% (was "Vancouver, BC, Canada}}" -- a compile error / warning source).
\author{Sharon Marfatia$^{3}$}
\affiliation{UBC AWS Cloud Innovation Centre, University of British Columbia,
Vancouver, BC, Canada}

\begin{abstract}
As large language models (LLMs) become embedded in quantum simulation
workflows (IDE copilots, notebook assistants, agentic pipelines),
evaluation must move beyond functional correctness to anticipate
and catch structured failures before they propagate through expensive
pipelines. We present a layered evaluation framework for materials-informed
Variational Quantum Eigensolver (VQE) circuit generation: (i)~a gatekeeper
screening rubric across seven physical and framework criteria; (ii)~a circuit
fidelity analysis comparing model outputs against analytical and
reference-implementation values for H$_2$/STO-3G/Jordan--Wigner/UCCSD, with
ansatz classification and gate-composition breakdown; and (iii)~design
entropy, a run-to-run behavioral consistency metric. We surface a taxonomy of
five distinct LLM failure modes (geometry hallucination, nonexistent API
usage, runtime integration failures, constraint violations, and
plausible-but-unverifiable output), each with distinct detectability profiles
and structural to the task rather than to any one model. A forensic audit of
the evaluation platform's own source code further establishes that two
apparent model failures originated in the harness through silent
fallback-template substitution, demonstrating that evaluation infrastructure
belongs inside the same trust boundary as the models it tests. Applied
across multiple foundation models on a Materials Project integrated pipeline,
the framework shows that gatekeeper-style validation is necessary, not
optional, for reliable deployment.
\end{abstract}

\maketitle

% ============================================================
% FIG 1 — keep as float before Introduction
\begin{figure*}[!t]
  \centering
  \includegraphics[width=0.97\textwidth,keepaspectratio]{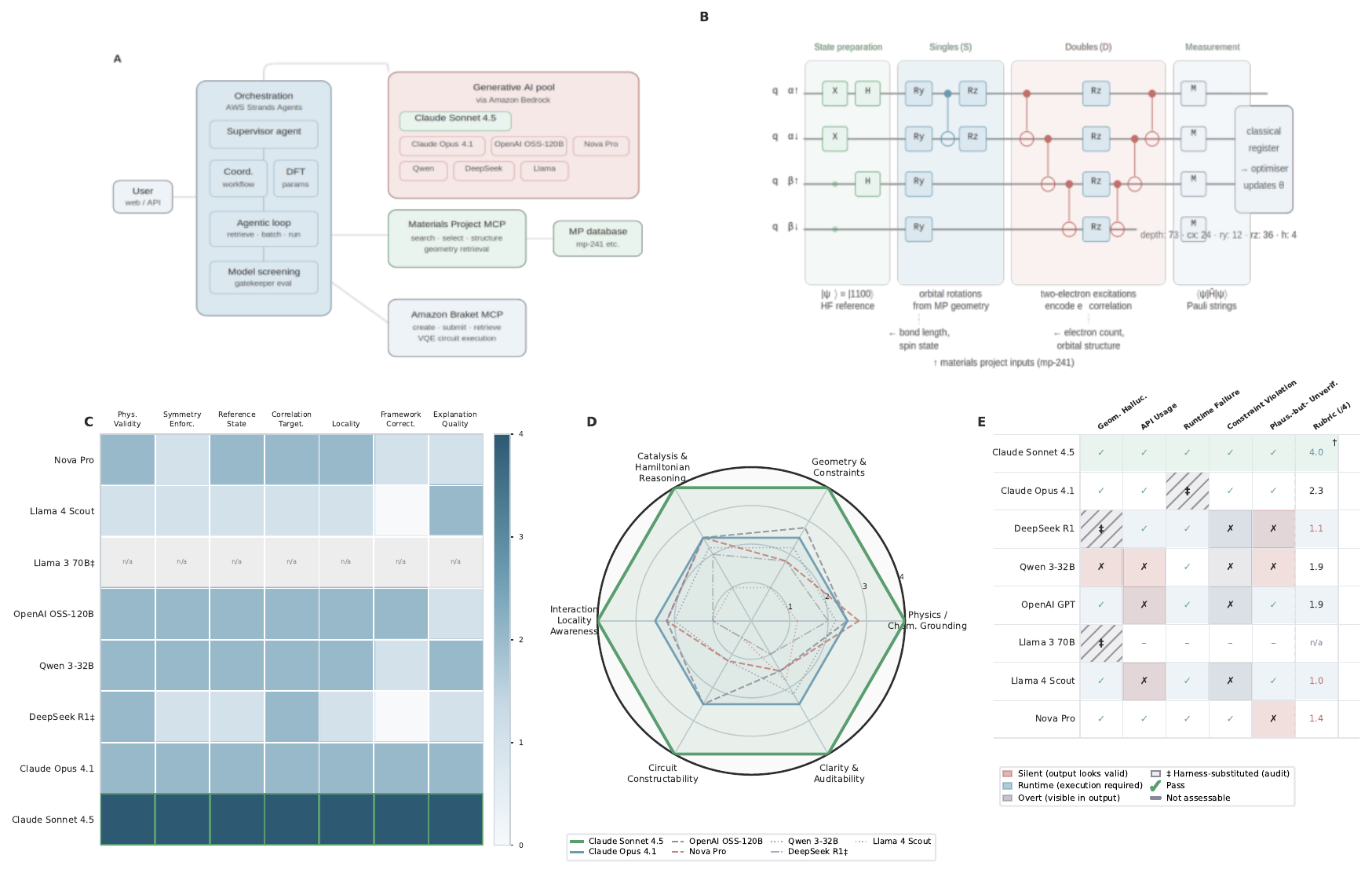}
  \caption{\textbf{System architecture, circuit structure, and screening overview.}
  \textbf{(A)}~Agentic pipeline schematic showing tool-mediated Materials Project
  retrieval via MCP server, multi-agent orchestration (Supervisor, Coordinator, DFT,
  and Structure agents), Amazon Braket circuit execution, and the generative AI model
  pool evaluated in this work.
  \textbf{(B)}~Schematic VQE circuit for H$_2$/STO-3G/Jordan--Wigner/UCCSD (mp-241)
  illustrating how Materials Project data encodes into gate structure: bond length and
  spin state determine the Hartree--Fock reference initialization; orbital count and
  electron number determine the singles and doubles excitation layers; the measurement
  layer feeds the classical optimizer. Gate counts from confirmed execution: depth~73,
  CX~$\times$24, R$_z$$\times$36, R$_y$$\times$12, H~$\times$4.
  \textbf{(C)}~Rubric heatmap (0--4 scale) across models and seven criteria. Deep blue
  indicates high scores; the green-bordered row marks the only model with confirmed
  executed output.
  \textbf{(D)}~Radar plot of capability profiles aggregated from rubric scores across
  six dimensions derived from the seven criteria. Capability profiles include
  model-reported metrics from Prompt~2~v3 and are not independently execution-verified.
  \textbf{(E)}~Failure taxonomy table showing, per model: ansatz class actually
  generated (from code inspection), Hartree--Fock reference usage, molecular identity,
  and mean rubric score (/4). Colour encodes failure detectability: pale red = silent
  (output looks valid), pale blue = runtime (requires execution to detect), grey =
  overt (visible in output). The Llama~3~70B and DeepSeek~R1 rows reflect
  harness-substituted template output rather than model-generated code (see
  Sec.~\ref{sec:results_p1}, Harness audit). \dag{}~Only model with confirmed
  executed output.}
  % TODO(figures): relabel the Llama 3 70B and DeepSeek R1 rows in panel E from
  % "Hubbard/TwoLocal" to e.g. "no extractable code -- harness template
  % substituted", and relabel the Opus 4.1 row from "runtime fail" to
  % "pipeline fail". Rescore the corresponding rows of panels C/D against the
  % models' actual emitted responses rather than the substituted templates.
  \label{fig:fig1}
\end{figure*}

\section{Introduction}
\label{sec:intro}

Variational Quantum Eigensolvers (VQEs)~\cite{Peruzzo2014,Tilly2022} are a central
approach for near-term quantum simulation of electronic structure and materials
properties. In practice, their use requires multiple brittle steps: selecting a
physically meaningful effective Hamiltonian, choosing an encoding, basis, and active
space, enforcing symmetries such as particle number and spin, and implementing
runnable circuits in rapidly evolving quantum SDKs~\cite{Javadi-Abhari2024,McArdle2020}.
The choice of ansatz family, whether chemistry-inspired (e.g., UCCSD and its adaptive
variants~\cite{Grimsley2019,Grimsley2023}) or hardware-efficient, critically determines both the
expressivity and trainability of the resulting circuit, with barren plateaus posing a
fundamental challenge to gradient-based optimization at
scale~\cite{Larocca2024,Fontana2024}.

These steps become more demanding for materials-adjacent workflows where inputs are
derived from external databases such as the Materials
Project~\cite{Jain2013,Horton2025} rather than directly from small molecular geometries.
State-of-the-art materials-informed VQE calculations, for example spin-defect
simulations combining DFT, quantum embedding theory, UCCSD ansatz construction,
postselection for unphysical states, and zero-noise extrapolation on real IBM
hardware~\cite{Huang2022}, require careful orchestration of multiple specialized codes
and expert judgment at each pipeline stage. Integrating database retrieval,
electronic-structure schema interpretation, and circuit synthesis into a coherent
automated pipeline remains a labor-intensive task that limits access to quantum
simulation tools for non-specialist researchers.

LLMs can potentially reduce this friction by generating code and circuit templates,
reasoning about physical constraints, and orchestrating multi-step workflows using
tools. Recent work has demonstrated increasingly capable LLM-based approaches to
quantum code generation. Domain-specific fine-tuning, exemplified by IBM's Qiskit
Code Assistant~\cite{Dupuis2025}, has demonstrated improved executable correctness on
SDK-level tasks. Benchmarks including Qiskit HumanEval~\cite{Vishwakarma2024} and
QCircuitBench~\cite{QCircuitBench2024} evaluate LLMs on executable quantum code across
a range of tasks. Fine-tuned LLMs have been shown to produce parameterized QAOA and
VQE circuits in OpenQASM~3.0 from datasets of 14{,}000 optimized
circuits~\cite{Jern2025}, and agentic reinforcement learning with quantum simulator
feedback can achieve near-perfect circuit validity on assembly-level
tasks~\cite{Yu2025}. Generative model approaches offer a complementary direction,
using transformer-based architectures to directly produce circuits optimized for target
Hamiltonians without explicit ansatz specification~\cite{GQE2024}. Autonomous agents
for quantum chemistry workflows~\cite{ZouElAgente2025,Cao2025} have demonstrated
multi-step task completion with adaptive error handling. Quantum simulation is
particularly well-suited to agentic AI because outputs are verifiable through simulation
or hardware execution~\cite{Dupuis2025}, enabling closed-loop optimization that is
difficult to achieve in other scientific domains.

However, the velocity of this development makes the absence of shared, physically
grounded evaluation infrastructure increasingly consequential. As LLM-assisted quantum
circuit generation moves from academic benchmarks toward deployment in research
pipelines---and integration deepens through IDE copilots, notebook assistants, and
automated workflow orchestrators---the key challenge is not only whether a model can
write syntactically valid code, but whether it can preserve physically meaningful
constraints, correctly integrate external database inputs, and produce circuits whose
structure reflects consistent design choices across repeated runs. Our experiments
reveal that failure in this domain is structured and diverse, and that different
failure modes carry fundamentally different risks: some are immediately visible,
others are silently incorrect and will pass superficial inspection while being
% REV: softened from asserted fact to stated expectation/hypothesis.
entirely wrong. Crucially, we expect that as model capabilities improve, output
plausibility will tend to increase faster than physical correctness---a conjecture
consistent with our observations here---which would make robust evaluation
infrastructure more rather than less important over time.

These failure modes are consistent with and extend the taxonomy of code hallucinations
identified in general-purpose code generation settings~\cite{Liu2024hallucination}, and
motivate a gatekeeper design that screens models before committing to expensive
materials-informed tasks. The primary contribution of this work is a layered evaluation
framework for materials-informed VQE circuit generation that is designed to be reusable
and model-agnostic: applicable equally to general-purpose and fine-tuned models, and
intended to serve as a shared evaluation foundation as the field continues to develop.
The framework proceeds in three stages---gatekeeper screening, structured failure
taxonomy with quantitative circuit fidelity analysis, and circuit-level behavioral
signatures---designed to characterize not only whether models succeed, but how and why
they fail. Compared to prior benchmarking work~\cite{Vishwakarma2024,QCircuitBench2024}
and LLM-driven circuit synthesis~\cite{Jern2025,Yu2025,GQE2024}, our framework is
distinguished by its focus on materials-database integration, multi-model behavioral
comparison under shared physical constraints, ansatz classification from first
principles, and design entropy as a run-to-run consistency metric.

% ============================================================
\section{System Overview}
\label{sec:system}

We study an agentic workflow deployed in a managed LLM environment with tool access.
The system separates concerns across modules: a retrieval layer for Materials Project
queries and schema interpretation~\cite{Jain2013,Ong2013}, a circuit synthesis layer
(Qiskit~\cite{Javadi-Abhari2024}) that generates Hamiltonian and active-space choices
and candidate ansatz families, and a reporting layer that emits structured outputs
enabling automated scoring. Tool access is mediated by MCP servers for Materials
Project retrieval and quantum workflow scaffolding, with prompts designed to support
reproducible evaluation across models.

A key design feature is the separation of a lightweight gatekeeper screening stage
from the heavier materials-informed ansatz task. This reflects a practical observation:
many models fail on fundamental physical or framework constraints before any materials
integration is attempted, making downstream task evaluation wasteful.

% ============================================================
\section{Evaluation Tasks}
\label{sec:tasks}

\paragraph*{Prompt 1: Baseline gatekeeper screening.}
Models are given a target molecule (H$_2$, STO-3G basis, Jordan--Wigner mapping,
Materials Project ID mp-241) and asked to emit executable Qiskit
code~\cite{Javadi-Abhari2024} implementing a UCCSD ansatz with a Hartree--Fock
reference state, printing circuit metrics with no optimizer loop or extraneous
commentary. Outputs are graded on a 0--4 rubric across seven criteria: Physical
Validity, Symmetry Enforcement, Reference State, Correlation Targeting, Locality,
% REV: "analytically correct reference values" -> reference values of Sec. IV;
% added explicit single-rater disclosure (cross-referenced in Limitations).
Framework Correctness, and Explanation Quality. Rubric scoring was performed by a
single author via direct code inspection and comparison against the reference values
defined in Sec.~\ref{sec:metrics}; the absence of independent raters is noted as a
limitation in Sec.~\ref{sec:discussion}.

\paragraph*{Prompt 2 (v3): Materials-informed ansatz generation.}
Models translate Materials Project fields~\cite{Jain2013} into a symmetry-preserving
VQE ansatz plan for crystalline silicon (mp-149), targeting low-energy electronic
structure near band edges under limited circuit depth. Ansatz families span
chemistry-inspired (UCCSD, pUCCJ) and hardware-efficient
designs~\cite{Tilly2022,Fedorov2022}. Metrics are model-reported; results characterize
design preferences rather than confirmed circuit properties.

% ============================================================
\section{Metrics}
\label{sec:metrics}

\paragraph*{M1: Screening rubric scores.}
Seven rubric criteria scored 0--4 provide a compact view of baseline model readiness.

\paragraph*{M2: Ansatz classification and circuit fidelity.}
We classify each model's output by the ansatz type actually instantiated in code,
independent of what was claimed. For models producing numeric circuit metrics, we
% REV: separated the genuinely analytic quantity (parameter count) from the
% implementation-dependent ones (depth, CX count), which depend on SDK version,
% decomposition level, and excitation ordering. Previously all three were
% described as "analytically correct," which overstates depth and gate count.
compare reported values against two classes of reference for
H$_2$/STO-3G/JW/UCCSD. The first is analytic: exactly 3~variational parameters
(1~double + 2~singles for a $(2e, 2o)$ active space after symmetry reduction),
derivable from electronic-structure first principles and independent of any software
implementation. The second is a reference implementation: circuit depth~73 and 24~CX
gates (Jordan--Wigner CNOT ladder), obtained from the Trotterized UCCSD ansatz at a
single level of \texttt{decompose()} in Qiskit~1.2.x with qiskit-nature~$\geq 0.7$.
Depth and gate counts are properties of a specific SDK version, decomposition level,
and excitation ordering rather than analytical constants, and we therefore treat them
as reference-implementation values. None of these references is derived from any
model output.

\paragraph*{M3: Design entropy.}
Normalized Shannon entropy over distinct design tuples (depth, two-qubit gate count,
parameter count) within each ansatz family across repeated runs. Higher entropy
indicates broader exploration; lower entropy indicates template-driven behavior. To
our knowledge, the application of this metric to characterize LLM behavioral
consistency in quantum circuit generation is novel, distinct from both token-level
diversity measures in code generation and circuit expressibility measures in the
hardware-efficient ansatz literature.
% REV: added small-sample caveat for the entropy estimator.
We note that plug-in Shannon entropy estimates carry a downward bias at small sample
sizes; given the modest number of repeated runs per family, the reported values should
be read as qualitative indicators of behavioral regime rather than precise quantities.

% ============================================================
% FIG 2 — full width, manually placed
\begin{figure*}[!t]
  \centering
  \includegraphics[width=0.97\textwidth,keepaspectratio]{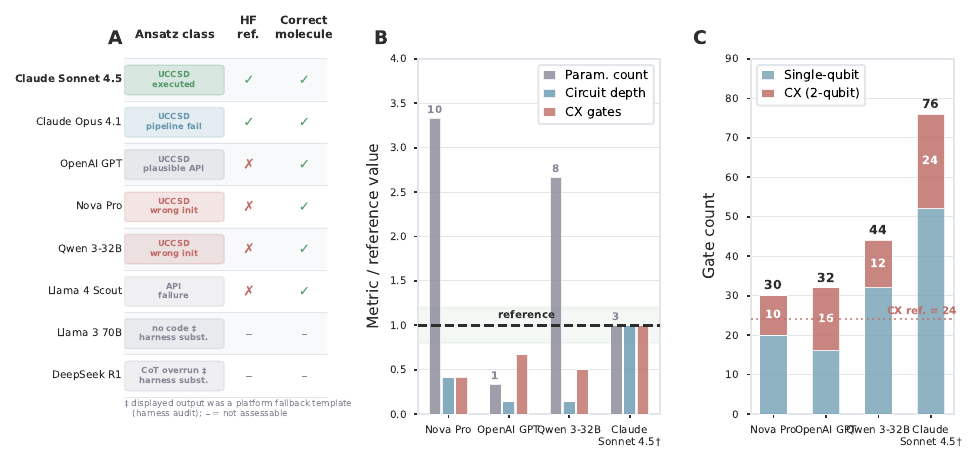}
  \caption{\textbf{Circuit fidelity and ansatz classification
  (H$_2$/STO-3G/JW/UCCSD, mp-241).}
  \textbf{(A)}~Ansatz classification per model from code inspection. Colour encodes
  failure detectability: pale red = silent (output looks valid), pale blue = runtime
  (requires execution to detect), grey = overt (visible in output). HF~ref.\ and
  correct molecule are binary checks. Entries for Llama~3~70B and DeepSeek~R1
  correspond to harness-substituted templates rather than model generations
  (Sec.~\ref{sec:results_p1}, Harness audit).
  % TODO(figures): relabel the Llama 3 70B and DeepSeek R1 rows accordingly and
  % the Opus 4.1 row to "pipeline fail"; optionally change the embedded
  % "analytical ref." line label in panel B to "reference".
  % REV: caption updated to distinguish analytic vs reference-implementation values.
  \textbf{(B)}~Variational parameter count, circuit depth, and CX gate count for
  models producing numeric output, normalised to the reference values (dashed line;
  params~$= 3$, analytic; depth~$= 73$ and CX~$= 24$, Qiskit~1.2.x reference
  implementation, see Sec.~\ref{sec:metrics}).
  \textbf{(C)}~Gate composition (CX vs.\ single-qubit). Dotted line: reference
  implementation CX count (24).}
  \label{fig:fig2}
\end{figure*}

% ============================================================
\section{Results}
\label{sec:results}

\subsection{Screening, Failure Taxonomy, and Circuit Fidelity}
\label{sec:results_p1}

Figure~\ref{fig:fig1} presents the complete screening evaluation. Panels~C and~D show
rubric scores and capability profiles, revealing substantial variability in constraint
handling. Panel~E summarises the failure taxonomy. Figure~\ref{fig:fig2} provides the
quantitative circuit fidelity analysis.

\textbf{Ansatz classification.}
Figure~\ref{fig:fig2}(A) classifies outputs by the ansatz type actually instantiated
in code. Only Claude Sonnet~4.5 produced confirmed executed UCCSD output. Claude
% REV: corrected attribution. The TypeError traceback originates in the
% evaluation platform's own response-handling code (html.unescape on a null
% response in app.py), not in the model-generated Qiskit code, which was never
% executed. Previously this was described as the model's code crashing.
Opus~4.1 generated structurally correct UCCSD code, but the surrounding pipeline
failed before that code could be executed: the platform's response-handling layer
raised a \texttt{TypeError} while post-processing a null return (traceable to a null
Materials Project retrieval upstream), a pipeline-level integration failure rather
than a defect in the generated circuit code. Two models
(Nova Pro, Qwen~3-32B) called \texttt{UCCSD()} with invalid constructor arguments.
OpenAI~GPT produced a plausible API call but with physically inconsistent parameter
counts. Llama~4~Scout invoked a nonexistent \texttt{Molecule.from\_database()} method.
% REV-FINAL: harness audit completed against the public repository; the hedged
% wording and ALTERNATIVE block have been replaced by the confirmed finding.
Llama~3~70B and DeepSeek~R1 appear in the platform transcripts as emitting
Hubbard/TwoLocal ans\"atze for CO rather than UCCSD for H$_2$. A forensic audit of
the platform source code (see \textbf{Harness audit} below) established that this
code was not generated by either model: both responses failed the platform's
code-extraction step (DeepSeek~R1 exhausted its token budget on chain-of-thought
before emitting a code block; Llama~3~70B produced no extractable code), and the
platform silently substituted its own pre-generated fallback template, populated
with an incorrectly resolved formula (CO) from the retrieval layer. We therefore
classify these two cases as constraint violation (DeepSeek~R1) and failure to emit
usable code (Llama~3~70B), each compounded by a harness-level silent substitution.
Only Claude Sonnet~4.5 and Opus~4.1 incorporated a Hartree--Fock
reference state.

\textbf{Harness audit.}
Byte-identical code blocks displayed under multiple models---a practical
impossibility for independent generations---prompted an audit of the platform's
public source code.\footnote{\url{https://github.com/UBC-CIC/Quantum-Matter-Institute-Streamlit-App},
file \texttt{models/base\_model.py}.} The audit established three facts. First, the
code blocks observed in the transcripts exist verbatim as f-string templates in the
platform repository: an auto-generated UCCSD template and a toy-Hubbard/TwoLocal
template. Second, the response-handling layer implements silent fallback
substitution: when fewer than 100 characters of code can be extracted from an LLM
response, the platform's own template is displayed in place of model output; the
substitution is recorded in internal logs but not surfaced in the interface. Third,
the molecular formula populating the template is resolved by the Materials Project
retrieval layer with hardcoded defaults, so the wrong-molecule content (CO)
originated in the pipeline's formula resolution rather than in any model. The audit
required no re-execution of the platform: source-code inspection combined with
transcript forensics (the DeepSeek~R1 reasoning trace correctly targets H$_2$ before
truncation) fully determines the attribution.

\textbf{Circuit metric fidelity.}
% REV: "analytically correct reference values" -> "reference values" with the
% analytic/implementation distinction made where each quantity is discussed.
Figure~\ref{fig:fig2}(B) compares reported circuit metrics against the reference
values of Sec.~\ref{sec:metrics}. H$_2$/STO-3G/JW/UCCSD has exactly 3~variational
parameters, derivable from first principles. Nova Pro reported 10~parameters
($+233\%$ error), Qwen~3-32B reported 8~($+167\%$), and OpenAI~GPT reported
1~($-67\%$). Models reporting depth~$\in\{10,30\}$ were not generating Trotterized
UCCSD, which produces depth~73 at a single level of \texttt{decompose()} in
Qiskit~1.2.x. Only Claude Sonnet~4.5 matched all three reference values.
The gate composition breakdown (Fig.~\ref{fig:fig2}C) shows that OpenAI~GPT's CX
fraction is more than twice the reference (0.67 vs.\ 0.32), while its parameter count
is undercounted---an internally inconsistent combination confirming the circuit type
was not UCCSD.

\textbf{Failure taxonomy.}
Beyond the quantitative analysis, our experiments surface five qualitatively distinct
failure modes~\cite{Liu2024hallucination}:

% REV-FINAL: instances confirmed harness-induced by source-code audit; the
% category is retained because wrong-molecule output is silently incorrect
% regardless of its proximate cause, and model-level geometry hallucination is
% documented elsewhere.
\textit{Geometry hallucination / wrong-molecule output}: syntactically valid
circuits with plausible metrics for the wrong molecule---detectable only by
inspecting the geometry string. In this study, the observed instances proved on
audit to be harness-induced: fallback-template substitution populated with an
incorrectly resolved formula, rather than model-generated code. The category
remains essential, both because wrong-molecule output is silently incorrect
whether it originates in the model or in the surrounding pipeline, and because
model-level hallucination of inputs is well documented in general code
generation~\cite{Liu2024hallucination}.

\textit{Nonexistent API usage}: calls to \texttt{Molecule.from\_database()} and
imports from nonexistent submodules produce structurally reasonable but non-runnable
code~\cite{Liu2024hallucination}.

% REV: corrected attribution to the pipeline post-processing layer; the analogy
% to runtime constraint enforcement is retained and arguably strengthened.
\textit{Runtime integration failure}: structurally correct UCCSD code from one model
was never executed because the evaluation pipeline itself crashed with a
\texttt{TypeError} while post-processing a null retrieval return---a software-level
analog of the unphysical-state problem in hardware VQE~\cite{Huang2022}. In both
cases, runtime constraint enforcement is required to catch failures that the
generated artifact alone would not reveal; notably, here the failing component was
the harness rather than the model, underscoring that evaluation infrastructure is
itself a source of silent failure.

\textit{Constraint violation}: models emitting verbose step-by-step reasoning or full
chain-of-thought when code-only output was specified, indicating unreliable instruction
following under strict output contracts. In one case, unterminated chain-of-thought
consumed the available token budget before any code was emitted, which in turn
triggered the harness substitution described above---an illustration of how model-level
and pipeline-level failures compound.

\textit{Plausible-but-unverifiable output}: summary outputs (e.g.,
``\texttt{parameters~=~10, depth~=~30}'') without runnable code, or metrics
inconsistent with the system's known physical properties.

\begin{figure}[!t]
  \centering
  \includegraphics[width=\columnwidth,keepaspectratio]{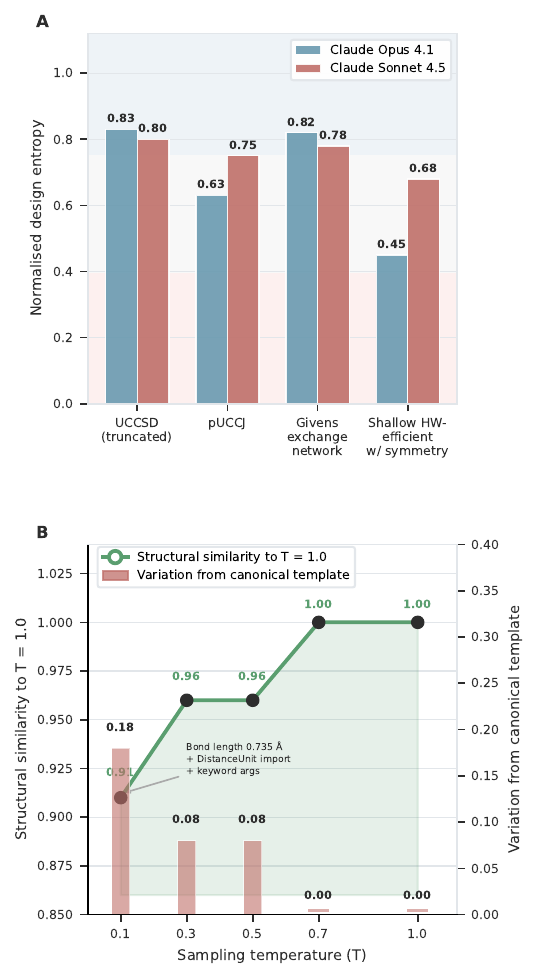}
  \caption{\textbf{Design entropy and temperature stability (model-reported metrics).}
  \textbf{(A)}~Normalized design entropy by model and ansatz family, computed from
  the empirical distribution of distinct (depth, two-qubit gate count, parameter count)
  tuples across repeated Prompt~2~v3 runs (crystalline silicon, mp-149). Higher entropy
  indicates broader exploration; lower entropy indicates template-driven generation.
  \textbf{(B)}~Temperature stability of Claude Sonnet~4.5 across $T \in
  \{0.1, 0.3, 0.5, 0.7, 1.0\}$ (Prompt~1, H$_2$/mp-241, $n = 5$ runs). Left axis:
  structural similarity to the $T = 1.0$ reference (green line). Right axis: variation
  from the canonical template (red bars). All outputs pass a functional checklist of
  8 criteria across the full temperature range. Circuit metrics in Panel~A are
  model-reported and not independently execution-verified.}
  \label{fig:fig3}
\end{figure}

% ============================================================
\subsection{Design Entropy and Behavioral Stability}
\label{sec:results_entropy}

Figure~\ref{fig:fig3} reports design entropy and temperature stability. Panel~A shows
normalized Shannon entropy by model and ansatz family for Prompt~2~v3 (crystalline
silicon). High entropy indicates broader exploration of the circuit design space across
repeated runs; low entropy indicates convergence to a small set of canonical templates.
These metrics characterize model behavior independently of whether individual circuit
values are execution-verified.

Panel~B reveals a qualitatively distinct behavioral finding: Claude Sonnet~4.5
maintains near-identical code structure and API choices across the full temperature
range $T \in \{0.1, 0.3, 0.5, 0.7, 1.0\}$, with structural similarity to the
$T = 1.0$ reference remaining at or above 0.96 for $T \geq 0.3$. The only measurable
deviation occurs at $T = 0.1$ (similarity~$= 0.91$), attributable to minor differences
in bond length specification (0.735~\AA\ vs.\ 0.74~\AA) and import style rather than
any change in circuit logic or API choice. Crucially, all five outputs pass a
functional checklist of eight criteria across the full temperature range, confirming
that the canonical UCCSD pipeline is preserved regardless of sampling temperature. This
temperature stability contrasts with findings in general code
generation~\cite{Liu2024hallucination}, where output diversity increases substantially
% REV: scoped the inductive-bias claim to this model/task and flagged it as
% preliminary (n=5 runs, one model, one prompt).
with sampling temperature, and is suggestive of a domain-specific inductive bias
toward physically grounded canonical designs for this model on this task. We
emphasize that this observation rests on $n = 5$ runs of a single model on a single
prompt and should be regarded as preliminary pending replication across models and
tasks.

% ============================================================
\section{Discussion}
\label{sec:discussion}

The results demonstrate that LLM failures in materials-informed VQE
workflows are neither random nor uniformly distributed: they are
structured, classifiable, and in several cases silently
incorrect~\cite{Liu2024hallucination}. A model that hallucinates
molecular geometry or invokes a nonexistent API method produces output
that may pass superficial inspection while being entirely wrong. This
has direct consequences for pipeline design. If, as we expect, output
% REV: softened from asserted fact to stated expectation (matching the Intro).
plausibility continues to improve faster than physical correctness as
model capabilities advance, silent failures will become harder to
detect over time.
Gatekeeper-style screening before committing to expensive
materials-informed tasks is therefore a necessary safeguard, not
merely an efficiency measure. Rubric criteria targeting geometry
correctness and API versioning are particularly high-value given the
frequency and severity of failures observed in these categories.
% REV-FINAL: the harness contributed two confirmed failures (post-processing
% crash; silent template substitution), both initially presenting as model
% failures and resolved only by forensic audit.
Notably, our own evaluation harness contributed two such failures: a
post-processing crash on a null response that initially presented as a
model-level error, and a silent fallback-template substitution that
initially presented as model-generated wrong-molecule code. Both passed
initial inspection and were correctly attributed only by forensic audit
of the platform source code, demonstrating that the gatekeeper principle
applies to evaluation infrastructure itself, not only to the models
under test.

The parameter count finding warrants specific attention. The correct
number of variational parameters for H$_2$/STO-3G/JW/UCCSD is
analytically derivable from first principles, yet multiple models
reported values inconsistent by large margins, indicating
pattern-matching to plausible-sounding outputs rather than reasoning
from the underlying electronic structure. For systems where the active
space permits analytical verification, parameter count provides a fast
and definitive check requiring no circuit execution. Extending such
checks to a broader library of reference cases spanning molecules,
basis sets, and mappings represents a tractable near-term investment
in evaluation infrastructure~\cite{Dupuis2025}.

Design entropy provides a complementary behavioral diagnostic. Rather
than constituting a performance ranking, entropy differences reflect
genuinely distinct use cases: high-entropy models are preferable when
surveying candidate ansatz families, while low-entropy models are
preferable when correctness and reproducibility are the priority. The
temperature stability result reinforces this framing---the observation
that canonical circuit structure and API choices are preserved across
the full sampling temperature range suggests that domain-specific
inductive biases toward physically grounded designs can be robust, a
useful property for reproducible agentic pipelines.

The framework is agnostic to model type and applies equally to
fine-tuned and general-purpose models, providing a shared evaluation
language across the landscape of emerging quantum software tools,
including domain-specific assistants~\cite{Dupuis2025} and
synthesis-oriented platforms.

\paragraph*{Limitations.}
Circuit metrics in Prompt~2~v3 are model-reported and some are
inconsistent with known physical properties. Materials Project
integration is tested primarily at the code-generation level, as most
models hardcode geometry rather than performing live database queries.
% REV: expanded limitations: single-rater rubric, harness-attribution audit,
% small-sample entropy bias.
Rubric scoring was performed by a single rater without blinding;
inter-rater reliability assessment is deferred to future work. For two
models, the displayed outputs were established by source-code audit to be
harness-substituted templates rather than model generations; the genuine
Prompt~1 behavior of these two models therefore remains uncharacterized
in this study, and re-evaluation in an isolated, harness-free setting is
deferred to future work. Sample sizes for
entropy analysis are modest---plug-in entropy estimates are downward
biased in this regime---and formal statistical comparisons are deferred
to future work. Future directions include execution-grounded
metrics~\cite{Fedorov2022}, quantum-verifiable
rewards~\cite{Dupuis2025}, live MP retrieval
validation~\cite{Horton2025}, and systematic ablations within a
standardized evaluation harness.

% ============================================================
\section{Conclusion}
\label{sec:conclusion}

We present a layered evaluation framework for materials-informed VQE circuit
generation, motivated by a challenge that will grow in practical importance as LLMs
become more deeply integrated into quantum simulation workflows: how to anticipate,
detect, and diagnose model failures before they propagate through expensive pipelines.
The framework---gatekeeper screening, a five-category failure taxonomy, quantitative
% REV: "analytical reference values" -> "analytical and reference-implementation
% values" for consistency with Sec. IV.
circuit fidelity analysis against analytical and reference-implementation values, and
design entropy as a behavioral consistency metric---is designed to be reusable and
model-agnostic. Its central contribution is not a ranking of current models, but a
vocabulary and methodology for characterizing failures that are structural properties
of this task class.

Geometry hallucination, nonexistent API usage, runtime integration failures, constraint
violations, and plausible-but-unverifiable outputs are distinct failure modes with
distinct detectability profiles and distinct consequences for downstream pipeline
integrity. These categories will remain relevant as the underlying models improve,
because the underlying challenge---grounding LLM-generated code in physical constraints
and external database schemas---does not disappear with scale.
% REV: added one sentence reflecting the harness-as-failure-source finding.
Our results further indicate that the evaluation harness itself belongs inside the
trust boundary: pipeline-level contamination and post-processing faults can masquerade
as model failures, and gatekeeper-style validation must therefore wrap the full
pipeline, not the model in isolation.

The circuit fidelity analysis demonstrates that parameter count error, derivable from
first principles, is the most accessible single diagnostic: a model reporting
physically inconsistent parameter counts has demonstrably failed to ground its output
in the underlying electronic structure, regardless of how syntactically plausible the
code appears. For practitioners, building a shared library of such analytically
verifiable reference cases---spanning molecules, basis sets, and mappings---is the
clearest near-term investment the community could make in evaluation infrastructure.
For tool designers, the gatekeeper pattern should be treated as a first-class
architectural component: as LLM integration deepens across the quantum software stack,
the cost of not having this infrastructure in place will increase. The evaluation
protocol presented here is intended as a reusable foundation for that effort,
supporting transparent and reproducible model assessment as agentic quantum tools
continue to develop.

% ============================================================
\begin{acknowledgments}
We thank collaborators at UBC SBQMI, the UBC Cloud Innovation Centre,
for discussions and support.
\end{acknowledgments}

\bibliographystyle{apsrev4-2}

% ============================================================
% APPENDIX
% ============================================================
\appendix

\section{AWS Platform Architecture and Implementation}
\label{app:aws}

This appendix describes the cloud infrastructure and multi-agent software
platform developed by the UBC Cloud Innovation Centre (CIC) in collaboration
with the Quantum Matter Institute (QMI) to support the benchmarking
experiments reported in this paper. The system was designed as a
proof-of-concept that exposes eight foundation models through a unified
interface, enabling reproducible, side-by-side evaluation of LLM-generated
quantum circuit code under controlled conditions. The architecture is
illustrated schematically in Fig.~\ref{fig:fig1}(A) of the main text.

\begin{figure*}[!t]
  \centering
  \includegraphics[width=0.97\textwidth,keepaspectratio]{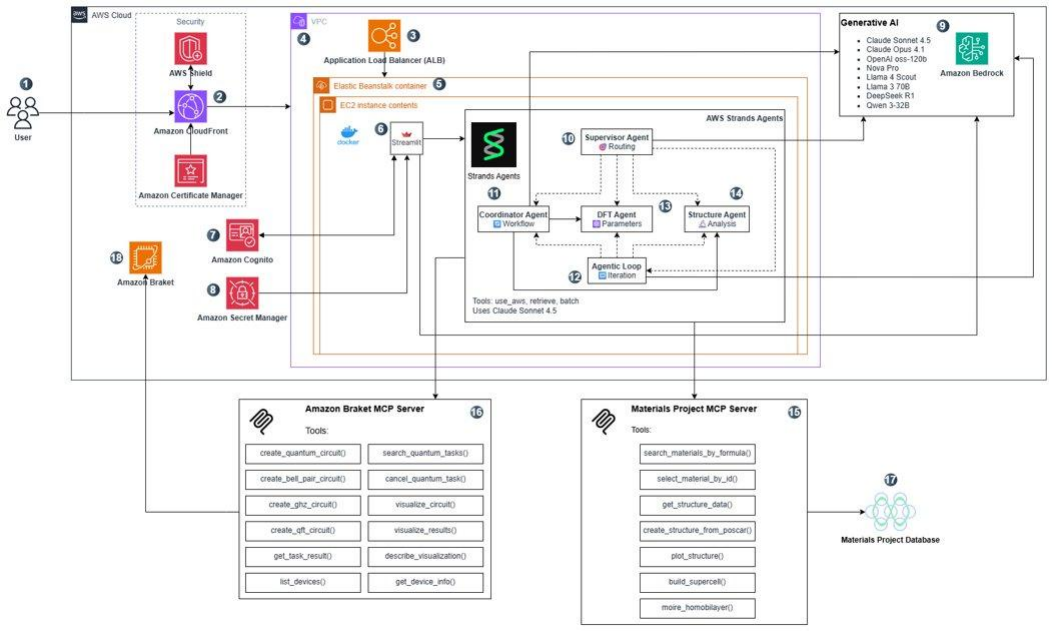}
  \caption{\textbf{AWS service architecture for the QMI multi-agent platform.}
  Numbered components correspond to the request flow described in the text.
  \textbf{(1--3)}~User traffic enters via Amazon CloudFront (with AWS Shield
  and Certificate Manager for security) and is routed through an Application
  Load Balancer to the Elastic Beanstalk container.
  \textbf{(4--6)}~Inside the EC2 instance, a Docker-containerised Streamlit
  application hosts the Strands Agents runtime.
  \textbf{(7--8)}~Amazon Cognito handles user authentication; AWS Secrets
  Manager provides secure credential storage for all API keys.
  \textbf{(9--10)}~The Supervisor Agent routes queries to Amazon Bedrock,
  which provides access to all eight foundation models evaluated in this
  work (Claude Sonnet~4.5, Claude Opus~4.1, OpenAI~OSS-120B, Nova Pro,
  Llama~4~Scout, Llama~3~70B, DeepSeek~R1, Qwen~3-32B).
  \textbf{(11--14)}~Specialised Strands agents --- Coordinator, DFT, Structure,
  and Agentic Loop --- orchestrate multi-step quantum materials workflows.
  \textbf{(15)}~The Materials Project MCP Server exposes tools for crystal
  structure retrieval, POSCAR generation, supercell construction, and
  moir\'{e} bilayer analysis against the Materials Project
  Database~\protect\cite{Jain2013}~\textbf{(17)}.
  \textbf{(16)}~The Amazon Braket MCP Server exposes tools for quantum circuit
  creation, submission, visualisation, and device management, connecting to
  Amazon Braket~\textbf{(18)} for simulator and hardware execution.}
  \label{fig:appfig}
\end{figure*}

\subsection{System Overview}
\label{app:aws:overview}

The platform is a containerised web application deployed on AWS Elastic
Beanstalk, accessible to researchers via a Streamlit front-end. Authentication
is handled by Amazon Cognito with JWT tokens; API credentials are stored in
AWS Secrets Manager and never exposed to the application layer. Global traffic
is routed through Amazon CloudFront (SSL termination and edge caching) to an
Application Load Balancer, which forwards requests to the Elastic Beanstalk
instance running the Docker container. All resources are isolated within a
Virtual Private Cloud (VPC).

\subsection{Multi-Agent Architecture}
\label{app:aws:agents}

Model inference and workflow orchestration are implemented using the AWS
Strands Agents SDK, an open-source, model-driven framework for building
production-ready AI agents. Three layers of agents are composed at runtime.
The Supervisor Agent acts as the top-level coordinator, receiving user queries,
selecting the appropriate sub-agent configuration, and aggregating results for
delivery to the front-end. The Coordinator and Agentic Loop layer orchestrates
multi-step workflows by dynamically instantiating specialised agents and
iterating until a converged result is obtained. At the specialised layer, the
DFT Agent handles density functional theory parameter extraction and Hamiltonian
construction, while the Structure Agent performs crystal structure parsing,
POSCAR file generation, and 3-D visualisation. All eight foundation models
(Claude Sonnet~4.5, Claude Opus~4.1, OpenAI~OSS-120B, Qwen~3-32B,
DeepSeek~R1, Nova Pro, Llama~4~Scout, Llama~3~70B) are served through Amazon
Bedrock via a unified API, making it straightforward to swap or extend the
model pool without changes to the agent logic.

\subsection{Model Context Protocol Server Integration}
\label{app:aws:mcp}

External scientific data sources are accessed through Model Context Protocol
(MCP) servers --- a standardised protocol that provides AI agents with secure,
structured access to external databases and tools. Two MCP servers were
deployed. The Materials Project MCP Server interfaces with the Materials
Project database~\cite{Jain2013} to retrieve crystallographic data, electronic
structure parameters, and geometry inputs (e.g.\ \texttt{mp-241} for H$_2$)
that are injected directly into circuit-generation prompts. The Amazon Braket
MCP Server provides the agent layer with real-time information about available
quantum devices and simulators, and supports circuit submission to the SV1
state-vector simulator and local simulators via the Amazon Braket SDK. Together,
the two MCP servers enable a dual quantum workflow: the Qiskit framework path
uses Materials Project geometry data to construct VQE circuits (as evaluated in
the main text), while the Braket framework path supports execution on AWS
quantum hardware and simulators with ASCII circuit visualisation.

\subsection{Deployment Configuration}
\label{app:aws:deploy}

All resources were deployed in the \texttt{ca\-central\-1} AWS region. The
application container runs on a \texttt{t3.large} EC2 instance managed by
Elastic Beanstalk, with Docker packaging the Streamlit application and all
Python dependencies for reproducible deployment.

\subsection{Post-Proof-of-Concept Development Roadmap}
\label{app:aws:roadmap}

Several infrastructure and capability improvements have been identified for
production deployment beyond the current proof-of-concept. On the data side,
integrating Amazon OpenSearch would enable retrieval-augmented generation over
a searchable corpus of quantum research papers and previously generated circuits,
accelerating research discovery. Evolving the stateless interface into a
persistent chat model would support iterative circuit refinement across sessions.
Connecting additional MCP data sources, including the NIST Chemistry WebBook
and the Crystallography Open Database, would broaden materials coverage beyond
the Materials Project. On the infrastructure side, transitioning from Elastic
Beanstalk to AWS CDK-deployed Lambda functions, or Amazon Bedrock AgentCore,
would improve cold-start latency, cost optimisation, and
infrastructure-as-code management. Resilience improvements, including connection
pooling for Bedrock clients and a circuit-breaker pattern for MCP server calls,
would prevent cascade failures under load. Two fixes identified directly by the
audit in Sec.~\ref{sec:results_p1} are also required for any production deployment:
the fallback code-substitution path in the response-handling layer must surface its
substitutions explicitly in the interface rather than only in internal logs, and
null model/retrieval responses must be handled before post-processing rather than
raising uncaught exceptions. The full source code for the platform
is publicly available at
\url{https://github.com/UBC-CIC/Quantum-Matter-Institute-Streamlit-App}.
% TODO(authors): insert the commit hash of the deployment used for the reported
% experiments in the sentence below (the repository has evolved since; e.g. the
% post-processing crash site has shifted line numbers between versions).
The experiments reported in the main text were performed against the deployment
corresponding to repository commit \texttt{<COMMIT-HASH>}.

\end{document}